# PERTURBATIONS IN A PLASMA

**Evangelos Chaliasos**

365 Thebes Street

GR-12241 Aegaleo

Athens, Greece

*Abstract*

The perturbations of a homogeneous non-relativistic two-component plasma are studied in the Coulomb gauge. Starting from the solution found [2] of the equations of electromagnetic self consistency in a plasma [1], we add small perturbations to all quantities involved, and we enter the perturbed quantities in the equations, keeping only the first order terms in the perturbations. Because the unperturbed quantities are solutions of the equations, they cancel each other, and we are left with a set of 12 linear equations for the 12 perturbations (unknown quantities). Then we solve this set of linearized equations, in the approximation of small ratio of the masses of electrons over those of ions, and under the assumption that the plasma remains homogeneous.

## 1. Introduction

In a two component plasma the following equations hold [1]:

$$\frac{c}{4\pi} \frac{\partial^2 A^l}{\partial x_i \partial x^i} = \sum_{+,-} j_\pm^l, \text{(Maxwell equations)} \quad (1.1)$$

$$\frac{1}{c}\left(A_i \frac{\partial n_\pm}{\partial x^k} - n_\pm \frac{\partial A_k}{\partial x^i}\right) u_\pm^i + \frac{c}{\lambda_\pm}\left(\frac{\partial n_\pm}{\partial x^k} - n_\pm u_\pm^i \frac{\partial u_{\pm k}}{\partial x^i}\right) = 0, \text{(equations of motion)} \quad (1.2)$$

$$\frac{\partial A^k}{\partial x^k} = 0, \text{(Lorentz gauge)} \quad (1.3)$$

$$\frac{\partial j_\pm^i}{\partial x^i} = 0. \text{(equation of continuity)} \quad (1.4)$$

Here $A^l$ is the four-potential ($\varphi$, **A**), and $j_\pm^l$ is the current four-vector, given by

$$j_\pm^i = c n_\pm u_\pm^i, \quad (1.5)$$

where $n_\pm$ is the scalar proper charge density and $u_\pm^i$ the four velocity field. $\lambda_\pm$ denotes the ratio of charge to mass, corresponding to one of the particles of each component. The subscript ± indicates which species the corresponding quantity is pertaining to: ions (+) or electrons (-).

In fact we have to consider only one of the equations of continuity (1.4), say the "plus" (+) one. The reason is that from Maxwell's equations (1.1) an equation of continuity for the total current $j_+^i + j_-^i$ can be extracted. Then the "minus" (-) equation of continuity can result if from the latter we subtract the "plus" (+) one. Note also that $j_+^i$ (and $j_-^i$ as well) has four (4) components, which are (because of eqn. (1.5)): $n_+$ and three of the $u_+^i$'s, say $u_+^\alpha$ ($\alpha$ = 1, 2, 3), because the four (4) components of $u_+^i$ are not all independent, since they are related by one (1) relation, namely

$$u_+^i u_{+i} = 1. \quad (1.6)$$

Thus we have totally: 4 Maxwell equations (1.1) plus 4×2(±) = 8 equations of motion (1.2) plus 1 Lorentz gauge (1.3) plus 1(+) equation of continuity (1.4) equal 14 equations, with: 4 $A^l$'s plus 4×2(±) = 8 $j_\pm^i$'s equal 12 unknowns. But, for reasons of independence, we have to skip the two (2)(±) 0-equations from the equations of motion (1.2), as explained in [2], so that we are finally left with 12 equations for 12 unknowns.

Note that we work with the potentials rather than the field intensities, and this is why a particular gauge (here the Lorentz gauge (1.3)) is required. Note also that the Maxwell equations (1.1) are valid only in this gauge according to this approach (see [1]).

For a homogeneous plasma, and in the non-relativistic approximation, the following solution to the above equations has been found [2]:

$$\vec{A} = \vec{a}\exp\{-i(\omega t - \vec{k}\cdot\vec{x})\}, \quad \phi = 0, \tag{1.7}$$

$$n_+ = -n_- \equiv n = \rho \text{ (say) } (= \text{const.}), \tag{1.8}$$

$$\vec{v}_\pm = -(\lambda_\pm/c)\vec{a}\exp\{-i(\omega t - \vec{k}\cdot\vec{x})\}, \tag{1.9}$$

or, instead of the latter, and because of (1.8),

$$\vec{j}_\pm = \mp\rho(\lambda_\pm/c)\vec{a}\exp\{-i(\omega t - \vec{k}\cdot\vec{x})\}. \tag{1.10}$$

Here **a** is a constant vector and (ω/c, **k**) a timelike four-vector, while $j_\pm^i$ = (cn$_\pm$ , **j**$_\pm$), with **j**$_\pm$ = n$_\pm$**v**$_\pm$. We observe that, in fact, there are: 3 components of **A** plus 1 (φ) plus 2 (n$_\pm$) plus 6 components (because of ±) of **v**$_\pm$ (or of **j**$_\pm$) equal 12 quantities as expected.

Note that the three (3) times two (2) (±) equal six (6) α-components (α = 1, 2, 3), plus the two (2) (±) 0-components equal 4×2(±) = 8 components of the four-velocities are given by

$$u_\pm^\alpha = -(\lambda_\pm/c^2)\vec{a}\exp\{-i(\omega t - \vec{k}\cdot\vec{x})\}, \tag{1.11}$$

$$u_\pm^0 = -(\lambda_\pm/c^2)C_\pm \cong 1, \tag{1.12}$$

or, instead of them, if we make use of the currents,

$$j_\pm^\alpha = -n_\pm(\lambda_\pm/c)\vec{a}\exp\{-i(\omega t - \vec{k}\cdot\vec{x})\}, \tag{1.13}$$

$$j_\pm^0 \equiv \pm\rho c = -n_\pm(\lambda_\pm/c)C_\pm \cong n_\pm c, \tag{1.14}$$

where C$_\pm$ are two constants appropriately adjusted.

## 2. The perturbation equations

If we add to the known solution, constituted by the 12 quantities $A^\alpha$, $\varphi$; $n_+$, $n_-$; $v_+^\alpha$, $v_-^\alpha$, the 12 small perturbations, denoted by the same symbols but with a dot over them, we demand that the resulting 12 perturbed quantities constitute also a solution. Thus, inserting them into equations (1.1), (1.2), (1.3) & (1.4), these equations must be satisfied as well. We obtain thus the following set of equations:

$$\{3\}: \frac{c}{4\pi} \frac{\partial^2 (A^\alpha + \dot{A}^\alpha)}{\partial x_i \partial x^i} = (n_+ + \dot{n}_+)(v_+^\alpha + \dot{v}_+^\alpha) + (n_- + \dot{n}_-)(v_-^\alpha + \dot{v}_-^\alpha) \quad (2.1)$$

$$\{1\}: \frac{c}{4\pi} \frac{\partial^2 (\phi + \dot{\phi})}{\partial x_i \partial x^i} = (n_+ + \dot{n}_+)c + (n_- + \dot{n}_-)c \quad (2.2)$$

$$\{3\}: \frac{1}{c}\left\{-(A_\alpha + \dot{A}_\alpha)\frac{\partial(n_+ + \dot{n}_+)}{\partial x^\beta} + (n_+ + \dot{n}_+)\frac{\partial(A_\beta + \dot{A}_\beta)}{\partial x^\alpha}\right\}\frac{(v_+^\alpha + \dot{v}_+^\alpha)}{c} +$$

$$+ \frac{1}{c}\left\{(\phi + \dot{\phi})\frac{\partial(n_+ + \dot{n}_+)}{\partial x^\beta} + (n_+ + \dot{n}_+)\frac{\partial(A_\beta + \dot{A}_\beta)}{c\partial t}\right\} \cdot 1 +$$

$$+ \frac{c}{\lambda_+}\left\{\frac{\partial(n_+ + \dot{n}_+)}{\partial x^\beta} - (n_+ + \dot{n}_+)\frac{(v_+^\alpha + \dot{v}_+^\alpha)}{c}\frac{\partial(-v_{+\beta} - \dot{v}_{+\beta})/c}{\partial x^\alpha} - (n_+ + \dot{n}_+) \cdot 1 \cdot \frac{\partial(-v_{+\beta} - \dot{v}_{+\beta})/c}{c\partial t}\right\} = 0 \quad (2.3)$$

$$\{3\}: \frac{1}{c}\left\{-(A_\alpha + \dot{A}_\alpha)\frac{\partial(n_- + \dot{n}_-)}{\partial x^\beta} + (n_- + \dot{n}_-)\frac{\partial(A_\beta + \dot{A}_\beta)}{\partial x^\alpha}\right\}\frac{(v_-^\alpha + \dot{v}_-^\alpha)}{c} +$$

$$+ \frac{1}{c}\left\{(\phi + \dot{\phi})\frac{\partial(n_- + \dot{n}_-)}{\partial x^\beta} + (n_- + \dot{n}_-)\frac{\partial(A_\beta + \dot{A}_\beta)}{c\partial t}\right\} \cdot 1 +$$

$$+ \frac{c}{\lambda_-}\left\{\frac{\partial(n_- + \dot{n}_-)}{\partial x^\beta} - (n_- + \dot{n}_-)\frac{(v_-^\alpha + \dot{v}_-^\alpha)}{c}\frac{\partial(-v_{-\beta} - \dot{v}_{-\beta})/c}{\partial x^\alpha} - (n_- + \dot{n}_-) \cdot 1 \cdot \frac{\partial(-v_{-\beta} - \dot{v}_{-\beta})/c}{c\partial t}\right\} = 0 \quad (2.4)$$

$$\{1\}: \frac{\partial(A^\alpha + \dot{A}^\alpha)}{\partial x^\alpha} + \frac{\partial(\phi + \dot{\phi})}{c\partial t} = 0 \quad (2.5)$$

$$\{1\}: \frac{\partial}{\partial x^\alpha}\left\{(n_+ + \dot{n}_+)(v_+^\alpha + \dot{v}_+^\alpha)\right\} + \frac{\partial}{c\partial t}\left\{(n_+ + \dot{n}_+)c\right\} = 0 \quad (2.6)$$

(The numbers in the curly brackets on the left indicate how many equations are on the right. We observe that we have in total 12 equations with 12 unknowns, namely the perturbations of the 12 quantities $A^\alpha$, $\varphi$; $n_+$, $n_-$; $v_+^\alpha$, $v_-^\alpha$)

Note that eqns. (2.1) are the three (3) spatial components of eqn. (1.1), eqn. (2.2) is the one (1) temporal component of the same eqn. (1.1), eqns.(2.3) are the three (3) spatial components of eqn. (1.2) with the plus (+) sign, eqns. (2.4) are similarly the three (3) spatial components of eqn. (1.2) with the minus (-) sign, eqn. (2.5) is eqn. (1.3), and eqn. (2.6) is eqn. (1.4) with the plus (+) sign: altogether 12 equations.

Expanding and keeping the terms up to the first order (in the perturbations) only (thus working in the linear approximation), and also omitting the zeroth order terms because they cancel out by means of the unperturbed equations which they satisfy, and simplifying, we are left with the following set of equations:

$$\{3\}: \frac{c}{4\pi} \frac{\partial^2 \dot{A}^\alpha}{\partial x_i \partial x^i} = n_+ \dot{v}_+^\alpha + v_+^\alpha \dot{n}_+ + n_- \dot{v}_-^\alpha + v_-^\alpha \dot{n}_- \qquad (2.7)$$

$$\{1\}: \frac{c}{4\pi} \frac{\partial^2 \dot{\phi}}{\partial x_i \partial x^i} = c\dot{n}_+ + c\dot{n}_- \qquad (2.8)$$

$$\{3\}: \frac{1}{c}\left\{-A_\alpha \frac{\partial \dot{n}_+}{\partial x^\beta} + n_+ \frac{\partial \dot{A}_\beta}{\partial x^\alpha} + \frac{\partial A_\beta}{\partial x^\alpha} \dot{n}_+\right\} \frac{v_+^\alpha}{c} +$$
$$+\frac{1}{c}\left\{n_+ \frac{\partial A_\beta}{\partial x^\alpha}\right\} \frac{\dot{v}_+^\alpha}{c} + \frac{1}{c}\left\{n_+ \frac{\partial \dot{A}_\beta}{c\partial t} + \frac{\partial A_\beta}{c\partial t} \dot{n}_+\right\} +$$
$$+\frac{c}{\lambda_+}\left\{\frac{\partial \dot{n}_+}{\partial x^\beta} + \left(n_+ \frac{\dot{v}_+^\alpha}{c} + \frac{v_+^\alpha}{c} \dot{n}_+\right)\frac{\partial v_{+\beta}}{c\partial x^\alpha} + n_+ \frac{v_+^\alpha}{c} \frac{\partial \dot{v}_{+\beta}}{c\partial x^\alpha} + n_+ \frac{\partial \dot{v}_{+\beta}}{c^2 \partial t} + \frac{\partial v_{+\beta}}{c^2 \partial t} \dot{n}_+\right\} = 0 \quad (2.9)$$

$$\{3\}: \frac{1}{c}\left\{-A_\alpha \frac{\partial \dot{n}_-}{\partial x^\beta} + n_- \frac{\partial \dot{A}_\beta}{\partial x^\alpha} + \frac{\partial A_\beta}{\partial x^\alpha} \dot{n}_-\right\} \frac{v_-^\alpha}{c} +$$
$$+\frac{1}{c}\left\{n_- \frac{\partial A_\beta}{\partial x^\alpha}\right\} \frac{\dot{v}_-^\alpha}{c} + \frac{1}{c}\left\{n_- \frac{\partial \dot{A}_\beta}{c\partial t} + \frac{\partial A_\beta}{c\partial t} \dot{n}_-\right\} +$$
$$+\frac{c}{\lambda_-}\left\{\frac{\partial \dot{n}_-}{\partial x^\beta} + \left(n_- \frac{\dot{v}_-^\alpha}{c} + \frac{v_-^\alpha}{c} \dot{n}_-\right)\frac{\partial v_{-\beta}}{c\partial x^\alpha} + n_- \frac{v_-^\alpha}{c} \frac{\partial \dot{v}_{-\beta}}{c\partial x^\alpha} + n_- \frac{\partial \dot{v}_{-\beta}}{c^2 \partial t} + \frac{\partial v_{-\beta}}{c^2 \partial t} \dot{n}_-\right\} = 0 \quad (2.10)$$

$$\{1\}: \frac{\partial \dot{A}^\alpha}{\partial x^\alpha} + \frac{\partial \dot{\phi}}{c\partial t} = 0 \qquad (2.11)$$

$$\{1\}: \frac{\partial}{\partial x^\alpha}\left(n_+ \dot{v}_+^\alpha + v_+^\alpha \dot{n}_+\right) + \frac{\partial}{c\partial t}(c\dot{n}_+) = 0 \qquad (2.12)$$

Finally, substituting the unperturbed quantities with their values given by eqns. (1.7), (1.8), (1.9), & (1.10), and after the simplifications, we are left with the following set of 12 perturbation equations:

$$\{3\}: \frac{c}{4\pi} \frac{\partial^2 \dot{A}^\alpha}{\partial x_i \partial x^i} = \rho\left(\dot{v}_+^\alpha - \dot{v}_-^\alpha\right) - a^\alpha \exp\cdot\left(\frac{\lambda_+}{c}\dot{n}_+ + \frac{\lambda_-}{c}\dot{n}_-\right) \quad (2.13)$$

$$\{1\}: \frac{c}{4\pi} \frac{\partial^2 \dot{\phi}}{\partial x_i \partial x^i} = c\dot{n}_+ + c\dot{n}_- \quad (2.14)$$

$$\{3\}: \frac{1}{c}\left\{-a_\alpha \exp\cdot\frac{\partial \dot{n}_+}{\partial x^\beta} + \rho\frac{\partial \dot{A}_\beta}{\partial x^\alpha}\right\}\left(-\frac{\lambda_+}{c^2}a^\alpha\right)\exp + \frac{1}{c}\rho\frac{\partial \dot{A}_\beta}{c\partial t} +$$

$$+ \frac{1}{\lambda_+}\left\{c\frac{\partial \dot{n}_+}{\partial x^\beta} - \rho\frac{\lambda_+}{c^2}a^\alpha \exp\cdot\frac{\partial \dot{v}_{+\beta}}{\partial x^\alpha} + \rho\frac{\partial \dot{v}_{+\beta}}{c\partial t}\right\} = 0 \quad (2.15)$$

$$\{3\}: \frac{1}{c}\left\{-a_\alpha \exp\cdot\frac{\partial \dot{n}_-}{\partial x^\beta} - \rho\frac{\partial \dot{A}_\beta}{\partial x^\alpha}\right\}\left(-\frac{\lambda_-}{c^2}a^\alpha\right)\exp - \frac{1}{c}\rho\frac{\partial \dot{A}_\beta}{c\partial t} +$$

$$+ \frac{1}{\lambda_-}\left\{c\frac{\partial \dot{n}_-}{\partial x^\beta} + \rho\frac{\lambda_-}{c^2}a^\alpha \exp\cdot\frac{\partial \dot{v}_{-\beta}}{\partial x^\alpha} - \rho\frac{\partial \dot{v}_{-\beta}}{c\partial t}\right\} = 0 \quad (2.16)$$

$$\{1\}: \frac{\partial \dot{A}^\alpha}{\partial x^\alpha} + \frac{\partial \dot{\phi}}{c\partial t} = 0 \quad (2.17)$$

$$\{1\}: \frac{\partial}{\partial x^\alpha}\left(\rho\dot{v}_+^\alpha - \frac{\lambda_+}{c}a^\alpha \exp\cdot\dot{n}_+\right) + \frac{\partial}{c\partial t}(c\dot{n}_+) = 0 \quad (2.18)$$

(Note: exp means exp{-i(ωt-**k**·**x**)}

Solving eqn. (2.14) for the perturbation of $n_-$, we find

$$\dot{n}_- = \frac{1}{4\pi}\frac{\partial^2 \dot{\phi}}{\partial x_i \partial x^i} - \dot{n}_+ . \quad (2.19)$$

Inserting this value of the perturbation of $n_-$ in eqn. (2.13), and solving the resulting equation for the perturbation of $n_+$, we find

$$\dot{n}_+ = \frac{c}{\lambda_+ - \lambda_-}\frac{-\frac{c}{4\pi}\frac{\partial^2 \dot{A}^\alpha}{\partial x_i \partial x^i} - \frac{\lambda_-}{4\pi c}\frac{\partial^2 \dot{\phi}}{\partial x_i \partial x^i}a^\alpha \exp + \rho\left(\dot{v}_+^\alpha - \dot{v}_-^\alpha\right)}{a^\alpha \exp}, \quad (2.20)$$

or

$$\dot{n}_+ a^\alpha \exp = \frac{c}{\lambda_+ - \lambda_-} \left\{ -\frac{c}{4\pi} D\dot{A}^\alpha - \frac{\lambda_-}{4\pi c} a^\alpha \exp D\dot{\phi} + \rho\left(\dot{v}_+{}^\alpha - \dot{v}_-{}^\alpha\right) \right\}, \qquad (2.21)$$

where D stands for the d′ Alembert operator. Similarly we find (inserting (2.21) in (2.19))

$$\dot{n}_- a^\alpha \exp = \frac{c}{\lambda_+ - \lambda_-} \left\{ \frac{c}{4\pi} D\dot{A}^\alpha + \frac{\lambda_+}{4\pi c} a^\alpha \exp D\dot{\phi} - \rho\left(\dot{v}_+{}^\alpha - \dot{v}_-{}^\alpha\right) \right\}. \qquad (2.22)$$

Also, because the combination

$$\frac{\lambda_+}{c} \dot{n}_+ + \frac{\lambda_-}{c} \dot{n}_- \qquad (2.23)$$

is independent of the index α in eqn. (2.13), we get from this equation the double relation

$$\frac{-\frac{c}{4\pi} D\dot{A}^1 + \rho\left(\dot{v}_+{}^1 - \dot{v}_-{}^1\right)}{A^1} = \frac{-\frac{c}{4\pi} D\dot{A}^2 + \rho\left(\dot{v}_+{}^2 - \dot{v}_-{}^2\right)}{A^2} = \frac{-\frac{c}{4\pi} D\dot{A}^3 + \rho\left(\dot{v}_+{}^3 - \dot{v}_-{}^3\right)}{A^3}. \qquad (2.24)$$

Thus, we can substitute the four equations (2.13) and (2.14) by the four equivalent equations (2.21) & (2.22), and (2.24).

We let eqn. (2.17) unchanged. But concerning eqn. (2.18), we can write it as

$$\text{div}(\rho\dot{\vec{v}}_+) + \frac{\partial}{c\partial t}(c\dot{n}_+) = \frac{\lambda_+}{c} \text{div}(\dot{n}_+ \vec{A}). \qquad (2.25)$$

Then, substituting the value of the perturbation of $n_+$ given by eqn. (2.21) in eqn (2.25), we get

$$\text{div}\left(\rho\dot{\vec{v}}_+\right) + \frac{c^2}{\lambda_+ - \lambda_-} \frac{\partial}{c\partial t} \frac{-\frac{c}{4\pi} D\dot{A}^1 + \rho\left(\dot{v}_+{}^1 - \dot{v}_-{}^1\right)}{A^1} - \frac{c}{4\pi} \frac{\lambda_-}{\lambda_+ - \lambda_-} \frac{\partial}{c\partial t} D\dot{\phi} =$$

$$\frac{\lambda_+}{\lambda_+ - \lambda_-} \text{div}\left[ \frac{-\frac{c}{4\pi} D\dot{A}^1 + \rho\left(\dot{v}_+{}^1 - \dot{v}_-{}^1\right)}{A^1} \vec{A} \right] - \frac{1}{4\pi c} \frac{\lambda_+ \lambda_-}{\lambda_+ - \lambda_-} \text{div}\left(\vec{A} D\dot{\phi}\right), \qquad (2.26)$$

which equation we can take instead of eqn. (2.18).

Concerning the 3 + 3 = 6 equations (2.15) & (2.16), we can write them respectively as

$$\frac{\lambda_+}{c^3} A_\alpha A^\alpha \frac{\partial \dot{n}_+}{\partial x^\beta} - \rho \frac{\lambda_+}{c^3} A^\alpha \frac{\partial \dot{A}_\beta}{\partial x^\alpha} + \frac{1}{c} \rho \frac{\partial \dot{A}_\beta}{c\partial t} + \frac{c}{\lambda_+} \frac{\partial \dot{n}_+}{\partial x^\beta} - \frac{\rho}{c^2} A^\alpha \frac{\partial \dot{v}_{+\beta}}{\partial x^\alpha} + \frac{\rho}{\lambda_+} \frac{\partial \dot{v}_{+\beta}}{c\partial t} = 0, \quad (2.27)$$

$$\frac{\lambda_-}{c^3} A_\alpha A^\alpha \frac{\partial \dot{n}_-}{\partial x^\beta} + \rho \frac{\lambda_-}{c^3} A^\alpha \frac{\partial \dot{A}_\beta}{\partial x^\alpha} - \frac{1}{c} \rho \frac{\partial \dot{A}_\beta}{c\partial t} + \frac{c}{\lambda_-} \frac{\partial \dot{n}_-}{\partial x^\beta} + \frac{\rho}{c^2} A^\alpha \frac{\partial \dot{v}_{-\beta}}{\partial x^\alpha} - \frac{\rho}{\lambda_-} \frac{\partial \dot{v}_{-\beta}}{c\partial t} = 0. \quad (2.28)$$

In order to simplify the problem, and since it is quite realistic to neglect the mass of the electrons as compared with the mass of the ions (nuclei), and having in mind application to a hydrogenic plasma (constituted only by protons and electrons), we consider the ratio of electron mass to proton mass to be negligible. Concretely, we set $\lambda_+/\lambda_- \equiv -\varepsilon \cong 0$, with $1 \gg \varepsilon > 0$, and $\lambda_- \equiv -e/m$, where e and m are the charge and the mass of the electron respectively. If we also use the fact that

$$\vec{v}_+ = -\frac{\lambda_+}{c} \vec{A} = -\varepsilon \left(-\frac{\lambda_-}{c} \vec{A}\right) = -\varepsilon \vec{v}_-, \quad (2.29)$$

so that

$$\dot{\vec{v}}_+ = -\varepsilon \dot{\vec{v}}_-, \quad (2.30)$$

then our set of perturbation equations becomes:

$$\{2\}: \frac{-\frac{c}{4\pi}D\dot{A}^1 + \rho(1+\varepsilon)\dot{v}_-^1}{A^1} = \frac{-\frac{c}{4\pi}D\dot{A}^2 + \rho(1+\varepsilon)\dot{v}_-^2}{A^2} =$$

$$= \frac{-\frac{c}{4\pi}D\dot{A}^3 + \rho(1+\varepsilon)\dot{v}_-^3}{A^3} \quad (2.31)$$

$$\{1\}: -\varepsilon \, \mathrm{div}\left(\rho\dot{\vec{v}}_-\right) - (1-\varepsilon)\frac{mc^2}{e}\frac{\partial}{c\partial t}\frac{-\frac{c}{4\pi}D\dot{A}^1 + \rho(1+\varepsilon)\dot{v}_-^1}{A^1} - \frac{c}{4\pi}(1-\varepsilon)\frac{\partial}{c\partial t}D\dot{\phi} =$$

$$= -\varepsilon\,(1-\varepsilon)\mathrm{div}\left[\frac{-\frac{c}{4\pi}D\dot{A}^1 + \rho(1+\varepsilon)\dot{v}_-^1}{A^1}\vec{A}\right] - \frac{1}{4\pi}\varepsilon\,(1-\varepsilon)\frac{e}{mc}\mathrm{div}\left(\vec{A}D\dot{\phi}\right) \quad (2.32)$$

$$\{1\}: \mathrm{div}\dot{\vec{A}} + \frac{\partial}{c\partial t}\dot{\phi} = 0 \quad (2.33)$$

$$\{3\}: \varepsilon\frac{e}{mc^3}A_\alpha A^\alpha \frac{\partial \dot{n}_+}{\partial x^\beta} - \rho\varepsilon\frac{e}{mc^3}A^\alpha \frac{\partial \dot{A}_\beta}{\partial x^\alpha} + \frac{\rho}{c}\frac{\partial \dot{A}_\beta}{c\partial t} +$$

$$+ \frac{1}{\varepsilon}\frac{mc}{e}\frac{\partial \dot{n}_+}{\partial x^\beta} + \varepsilon\frac{\rho}{c^2}A^\alpha \frac{\partial \dot{v}_{-\beta}}{\partial x^\alpha} - \rho\frac{m}{e}\frac{\partial \dot{v}_{-\beta}}{c\partial t} = 0 \quad (2.34)$$

$$\{3\}: -\frac{e}{mc^3}A_\alpha A^\alpha \frac{\partial \dot{n}_-}{\partial x^\beta} - \rho\frac{e}{mc^3}A^\alpha \frac{\partial \dot{A}_\beta}{\partial x^\alpha} - \frac{\rho}{c}\frac{\partial \dot{A}_\beta}{c\partial t} -$$

$$- \frac{mc}{e}\frac{\partial \dot{n}_-}{\partial x^\beta} + \frac{\rho}{c^2}A^\alpha \frac{\partial \dot{v}_{-\beta}}{\partial x^\alpha} + \rho\frac{m}{e}\frac{\partial \dot{v}_{-\beta}}{c\partial t} = 0 \quad (2.35)$$

$$\{1\}: \dot{n}_+ = (1-\varepsilon)\frac{mc}{e}\frac{+\frac{c}{4\pi}D\dot{A}^1 + \frac{e}{4\pi mc}A^1 D\dot{\phi} - \rho(1+\varepsilon)\dot{v}_-^1}{A^1} \quad (2.36)$$

$$\{1\}: \dot{n}_- = (1-\varepsilon)\frac{mc}{e}\frac{-\frac{c}{4\pi}D\dot{A}^1 + \varepsilon\frac{e}{4\pi mc}A^1 D\dot{\phi} + \rho(1+\varepsilon)\dot{v}_-^1}{A^1} \quad (2.37)$$

(Here: eqn. (2.31) is eqn. (2.24), eqn. (2.32) comes from eqn. (2.26), eqn. (2.33) is eqn. (2.17), eqn. (2.34) respectively (2.35) come from eqn. (2.27) respectively (2.28), and eqn. (2.36) respectively (2.37) come from eqn. (2.21) respectively (2.22)).

### 3. Solution of the perturbation equations

First of all, if we write down eqn. (20) of [2] for the perturbed plasma, we will have

$$\frac{c^4}{\lambda_+^2} - (\dot{\phi} + C'_+)^2 = \frac{c^4}{\lambda_-^2} - (\dot{\phi} + C'_-)^2, \qquad (3.1)$$

where $C'_\pm$ are two constants, the primes pertaining to the perturbed solution. Thus we see that the perturbation of the scalar potential must be constant.

Then, eqn. (1.1) gives, because of eqn. (1.5),
$$0 = j'^0_+ + j'^0_- = cn'_+ u'^0_+ + cn'_- u'^0_-. \qquad (3.2)$$

Introducing $u'^0_\pm$ from
$$u'^0_\pm = -\frac{\lambda_\pm}{c^2}(\dot{\phi} + C'_\pm), \qquad (3.3)$$

which is eqn. (18) of [2] written for the perturbed state, into eqn. (3.2), expanding the resulted equation and making use of eqn. (22) of [2], we find, keeping only the terms up to the first order,
$$\lambda_+ n_+ (\dot{\phi} + \dot{C}_+) + \lambda_- n_- (\dot{\phi} + \dot{C}_-) + \lambda_+ \dot{n}_+ C_+ + \lambda_- \dot{n}_- C_- = 0. \qquad (3.4)$$

Substituting $n_+ = -n_- = \rho$, with the result that also the same equation holds for the corresponding perturbations (see below), in (3.4) and making use of eqn. (40) of [2], we finally have
$$\lambda_+ (\dot{\phi} + \dot{C}_+) = \lambda_- (\dot{\phi} + \dot{C}_-). \qquad (3.5)$$

Now, without loss of the generality, we can set
$$\dot{\phi} = 0, \qquad (3.6)$$

thus working in a special case of the Lorentz gauge, in the Coulomb gauge. Then eqns. (3.1) and (3.5) become respectively
$$\frac{c^4}{\lambda_+^2} - (C_+ + \dot{C}_+)^2 = \frac{c^4}{\lambda_-^2}(C_- + \dot{C}_-)^2 \qquad (3.7)$$

and
$$\lambda_+ \dot{C}_+ = \lambda_- \dot{C}_-. \qquad (3.8)$$

But we know $C_\pm$. It is given by eqn. (40) of [2]. Thus eqn. (3.7) becomes

$$\frac{c^4}{\lambda_+^2} - \left(\dot{C}_+ - \frac{c^2}{\lambda_+}\right)^2 = \frac{c^4}{\lambda_-^2} - \left(\dot{C}_- - \frac{c^2}{\lambda_-}\right)^2. \tag{3.9}$$

From eqns. (3.8) & (3.9) we can determine the perturbations in the constants.

From eqn. (34) of [2], written down for the perturbed state, namely
$$dn'_\pm/ds = 0, \tag{3.10}$$

we find $n'_\pm$ = constant. But then, because there are no sources and sinks, we obtain
$$\dot{n}_\pm = 0. \tag{3.11}$$

We have found so far the perturbations in the temporal components of $A^i$ & $u_\pm^i$ ($\varphi$ & $n_\pm$). Now, we observe that eqns. (2.36) & (2.37) coincide, giving, to the first order in $\varepsilon$,
$$D\dot{A}^1 = (1+\varepsilon)\frac{4\pi\rho}{c}\dot{v}_-^1. \tag{3.12}$$

Also, eqns. (2.31), because of eqn. (3.12), give
$$D\dot{A}^\alpha = (1+\varepsilon)\frac{4\pi\rho}{c}\dot{v}_-^\alpha. \tag{3.13}$$

Eqns. (3.13) correspond to the Maxwell equations. We observe that they contain eqn. (3.12). We have also from eqn. (2.33)
$$\text{div}\dot{\bar{A}} = 0, \tag{3.14}$$

which equation corresponds to the Coulomb gauge, and from eqn. (2.32)
$$\text{div}\dot{\bar{v}}_- = 0, \tag{3.15}$$

which corresponds to the equation of continuity. The remaining equations are the equations of motion (2.34) & (2.35), which equations we will examine at the end.

But we know that
$$\bar{v}_- = -\frac{\lambda_-}{c}\bar{A}, \tag{3.16}$$

which equation results in
$$\dot{\bar{v}}_- = \frac{e}{mc}\dot{\bar{A}}. \tag{3.17}$$

Thus we see that the continuity equation (3.15) gives simply the gauge equation (3.14), which we will examine at the end, before the equations of motion. Thus we are left with the three (3) Maxwell equations (3.13). Substituting eqn. (3.17) in it results in being left with the three (3) equations with three (3) unknowns (the perturbations of the vector potential $A^\alpha$):

$$D\dot{\vec{A}} = K\dot{\vec{A}}, \qquad (3.18)$$

with the constant K given by

$$K = (1+\varepsilon)4\pi\,\rho\,\frac{e}{mc^2}. \qquad (3.19)$$

We know that

$$\vec{A} = \vec{a}\exp\{-i(\omega t - \vec{k}\cdot\vec{x})\}, \qquad (3.20)$$

resulted as a solution (eigenvector) of the eigenvalue equation of the operator D

$$D\vec{A} = C\vec{A}, \qquad (3.21)$$

with the constant C (eigenvalue) given by

$$C = \frac{4\pi}{c^2}(\lambda_+ - \lambda_-)\rho \qquad (3.22)$$

(cf. sec. 5 of [2]). If we take the variation of **A** from equation (3.20), we obtain

$$\dot{\vec{A}} = [\dot{\vec{a}} - i\vec{a}(\dot{\omega}t - \dot{\vec{k}}\cdot\vec{x})]\exp\{-i(\omega t - \vec{k}\cdot\vec{x})\}. \qquad (3.23)$$

On the other hand the perturbation of **A** must be a solution (eigenvector) of the same eigenvalue equation (for the operator D) (3.18), *corresponding to the same eigenvalue,* since, as we can easily find,

$$K = C. \qquad (3.23')$$

But

$$C = \frac{\omega^2}{c^2} - \vec{k}^2, \qquad (3.24)$$

while

$$K = \frac{\omega'^2}{c^2} - \vec{k}'^2, \qquad (3.24')$$

where ω´, **k**´ (and **a**´) are defined through the equation (analogous to eqn (3.20))

$$\dot{\vec{A}} = \vec{a}'\exp\{-i(\omega't - \vec{k}'\cdot\vec{x})\} \qquad (3.25)$$

(solution of (3.18)). Thus, comparing eqn. (325) with eqn. (3.23), we see that we must have

$$\omega' = \omega, \quad \vec{k}' = \vec{k}, \qquad (3.26)$$

that is

$$\dot{\omega} = 0, \quad \dot{\vec{k}} = 0, \qquad (3.27)$$

and

$$\vec{a}' = \dot{\vec{a}}. \qquad (3.28)$$

This happens because **A** and its perturbation are exactly eigenvectors of D *corresponding to the same eigenvalue* (K = C). Thus we are finally left with a solution for the perturbation of **A**

$$\dot{\vec{A}} = \dot{\vec{a}} \exp\{-i(\omega t - \vec{k} \cdot \vec{x})\}. \qquad (3.29)$$

We have to observe here that the perturbation of **a** *cannot be taken arbitrarily.* In fact it must be taken in such a way that the gauge eqn. (3.14) is satisfied. Thus, we can easily find that the condition

$$\dot{\vec{a}} \cdot \vec{k} = 0 \qquad (3.30)$$

must be fulfilled, that is it has to be orthogonal to **k**.

Finally, concerning the equations of motion (2.34) & (2.35), if we take in mind eqn (3.17), it is trivial for someone to see that they are satisfied identically.

Thus, the problem of the solution of the perturbation equations, in the special case of an *incompressible* plasma fluid, has been solved. The solution consists of only taking the unperturbed solution and substituting **a** by its perturbation. We can admire here the triviality of the solution, despite the complexity of the perturbation equations! Pertaining to the stability of the unperturbed solution, it is obvious that it is not unstable, since the perturbation is bounded.